\documentclass[runningheads]{preprint}

\usepackage[numbers, compress]{natbib}

\usepackage[utf8]{inputenc} 
\usepackage[T1]{fontenc}    
\usepackage{hyperref}       
\usepackage{url}            
\usepackage{booktabs}       
\usepackage{amsfonts}       
\usepackage{nicefrac}       
\usepackage{microtype}      
\usepackage{xcolor}         

\usepackage{amsmath}
\usepackage{graphicx}
\usepackage{wrapfig}
\usepackage{todonotes}

\title{Learning Hybrid Biophysical Neuron Models\\ with Neural ODEs}

\author{Jonas Beck\inst{1,2,3} \and
Michael Deistler\inst{2,3,4} \and
Dóra Viktória Molnár\inst{2,3} \and
Jakob H. Macke\inst{2,3,5} \and
Philipp Berens\inst{1,2}}

\institute{Hertie Institute for AI in Brain Health, Tübingen, Germany\\ \and
Tübingen AI Center, University of Tübingen, Tübingen, Germany \\ \and
Machine Learning in Science, Excellence Cluster Machine Learning, University of Tübingen, Tübingen, Germany\\ \and
Max Planck Institute for Biological Intelligence, München, Germany\\ \and
Department Empirical Inference, Max Planck Institute for Intelligent Systems, Tübingen, Germany\\
\email{\{jonas.beck, philipp.berens\}@uni-tuebingen.de}\\
}

\begin{document}

\maketitle

\begin{abstract}
Biophysical neuron models link measurements of neural activity to underlying cellular mechanisms. 
Yet, a central challenge is that the kinetics of many ion channels are poorly characterized, and practical simplifications---omitting channels or reducing morphological detail---introduce systematic gaps between model and biology. 
Bridging these gaps requires approaches that can flexibly discover unmodeled dynamics while preserving mechanistic interpretability.
Here, we introduce a hybrid modeling framework that embeds neural ordinary differential equations into conductance-based biophysical models to capture unknown currents or mis-specified channel kinetics. 
By parameterizing the neural ODE in terms of voltage-dependent steady-state and time-constant functions, we recover interpretable gating dynamics directly from voltage recordings without assuming a functional form.
We show that the hybrid model fits the gating kinetics of 2400 ion channel models and recovers unknown gating dynamics from single current-clamp recordings, generalizing to out-of-distribution stimulus regimes under realistic inputs and parameter misspecification. 
We also use our method to reduce a multicompartment model of a cortical neuron into a single-compartment hybrid model with a learned axial current, yielding up to an order of magnitude lower computational cost. 
Together, our results establish a plug-and-play framework for selectively replacing unknown components of conductance-based models with neural ODEs while preserving their mechanistic structure.
\end{abstract}

\section{Introduction}
Biophysical models connect observable neural activity and the underlying cellular mechanisms. Hodgkin--Huxley-type models \cite{hodgkin1952quantitative} describe how cellular voltage fluctuations---and in particular action potentials---are generated by ionic currents flowing through different channels embedded in a cell's membrane. Depending on the type of channel and the membrane voltage, channels open and close at characteristic rates.
The gating functions that determine these rates are typically established through labor-intensive voltage-clamp experiments, and their kinetics are fit using pre-specified parametric equations which can lack flexibility \cite{destexhe2000which, podlaski2017mapping, chintaluri2025omnimodel}. 
For many channels, good descriptions remain unknown, leaving some dynamics of the observed cellular activity unaccounted for. Finally, simplifications in the modeled neural morphology may lead to further gaps between the computational model and the biological system it aims to describe. Thus, these gaps arise from two distinct sources: unknown or poorly characterized channel kinetics, and morphological simplifications that omit dendritic contributions to somatic dynamics. 

\begin{figure}
    \centering
    \includegraphics[width=1.0\linewidth]{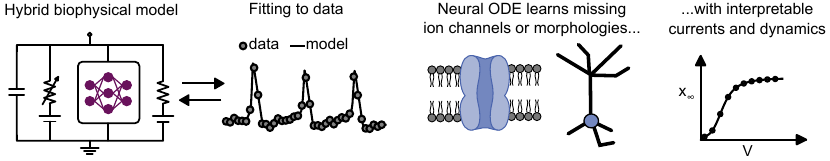}
    \caption{\textbf{Hybrid biophysical models embed neural ODEs into biophysical models to obtain flexible, efficient, and interpretable models of neural dynamics}. Upon fitting to data with gradient descent, the neural ODE learns to compensate for missing ion channels or morphological simplifications, yielding accurate and computationally efficient biophysical models. The resulting dynamics and currents are interpretable, which enables direct comparisons to experimental measurements and can inspire biophysical mechanisms.}
    \label{fig:graphical_abstract}
\end{figure}

In both cases, the challenge is to infer missing dynamics from observed voltage recordings without discarding the mechanistic structure that makes biophysical models interpretable. 
One approach has been to forego the mechanistic structure altogether and to learn dynamics directly from data using flexible sequence models such as recurrent neural networks \cite{sussillo2015neural, Pandarinath_2018}, state-space models \cite{gu2022efficiently}, or neural ordinary differential equations (neural ODEs) \cite{chen2018neural}, which can capture complex temporal dynamics without pre-specifying the governing equations. End-to-end approaches have shown that neural networks can accurately reproduce neuronal behavior, from operator learning \cite{centofanti2024learning} to systematic evaluations of physics-informed neural networks and neural ODEs \cite{matzakos2026comparing}. However, what these approaches gain in flexibility, they sacrifice in mechanistic transparency: While they can accurately model data, extracting biophysical insights can be difficult. 

Hybrid approaches that embed neural networks within mechanistic differential equations offer a way to navigate this tension by preserving known structure while learning unknown components from data. Universal differential equations \cite{rackauckas2020universal} have formalized this idea and it has since been applied, for example, to learn unknown 
subgrid-scale parameterizations in ocean turbulence models 
\cite{ramadhan2020capturing}, glacier ice flow rheology 
\cite{bolibar2023universal}, and bio-reactor dynamics 
\cite{strouwen2026bayesian}, demonstrating that neural networks can learn physically meaningful components across diverse scientific domains. While initial work has explored this direction in neuroscience \cite{estienne2023hybrid, ghanem2024learning, burghi2024recurrent, elgazzar2025universal}, the potential of such approaches for discovering biophysical mechanisms from electrophysiological data has not been fully realized, especially in light of newly developed differentiable biophysical simulators \cite{deistler2025jaxley}.
%

%
Here, we introduce a hybrid modeling framework that integrates neural ODEs into biophysical neuron models to recover the dynamics of unaccounted-for ionic and axial currents. The key idea is that any current term in a conductance-based model can be selectively replaced by a neural ODE component while the rest of the model remains mechanistic. When applied to ionic currents, the neural network outputs gating dynamics in terms of voltage-dependent steady-states and time constants, making the recovered dynamics inherently interpretable. 

We first show that the neural ODE formulation can represent the gating kinetics of 2400 ion channel models, matching expert-parameterized models while being substantially easier to train. We then demonstrate that our approach recovers the gating dynamics of unknown channels from voltage recordings alone, even under observation noise and with mis-tuned biophysical parameters, and generalizes to out-of-distribution stimulus regimes. When applied to axial currents, the same framework learns effective current terms and their latent dynamics from somatic recordings of morphologically detailed neurons, enabling a single-compartment model to faithfully reproduce multi-compartment behavior, directly addressing the gap introduced by morphological simplification. This yields both a computationally lighter model and interpretable latent variables that capture the dendritic contributions shaping somatic activity.

\section{Related Work}\label{sec:related_work}
Ion channel omnimodels parameterize the steady-state and time constant of ion channels in a fixed (but flexible) functional form \cite{chintaluri2025omnimodel}. These models can, in principle, describe a wide range of ion channel kinetics, but require elaborate fitting procedures even when voltage-clamp data is available. Our neural ODE formulation matches their performance with a generic architecture that is simpler to train.

At the other extreme, purely data-driven approaches \cite{centofanti2024learning, kainth_2025, matzakos2026comparing} can accurately reproduce neuronal dynamics but discard the mechanistic structure of the biophysical model entirely. In contrast, our framework constrains the learned components within the conductance-based formalism. As such, the 
recovered gating functions can be directly compared to experimental measurements, used in standard simulators, or distilled into closed-form expressions via symbolic regression \cite{cranmer2023interpretable}.

Several hybrid approaches have been explored in the context of neuroscience. \citet{ghanem2024learning} applied physics-informed neural ODEs to learn unknown dynamics but assumed that gating variables are directly observed. \citet{burghi2024recurrent} learned effective membrane currents using recurrent mechanistic models, but did not recover the underlying gating dynamics in terms of $x_\infty(V)$ and $\tau_x(V)$, limiting the biophysical interpretability of the learned representations. \citet{lei2021neural} used neural network differential equations to model ion channel kinetics in a hybrid fashion for the hERG potassium channel, but focused on cardiac applications and did not address partial observability or multicompartment reduction. The closest prior work is \citet{estienne2023hybrid}, who also learned Hodgkin--Huxley rate functions from current-clamp data. However, their approach uses a fixed-step forward Euler discretization rather than integrating through an ODE solver, entangling discretization error with the learned dynamics. It also employs minimal hand-crafted networks with built-in monotonicity constraints, limiting expressiveness beyond standard parametric forms, and does not allow co-optimization of biophysical parameters. 
Finally, none of the approaches above have addressed learning effective axial currents to reduce multicompartment models to single-compartment surrogates, which we demonstrate as a second application of our framework.

\section{Methods}

\subsection{Biophysical Models}\label{sec:biophysical_models}
Biophysical neuron models provide a mathematical description of how voltage fluctuations  are generated in response to an external stimulus dependent on the dynamics of ion channels in the neuron's membrane. For a compartment $n$ of a biophysical neuron model, the evolution of the membrane potential can be described by the ordinary differential equation
\begin{equation}
    C_n \frac{\text{d}V_n}{\text{d}t} = I_{n}^{mem} + I_{n}^{axial}, \label{eq:multicomp_formalism}
\end{equation}
where $C_n$ is the membrane capacitance, $V_n$ is the membrane potential and $I_{n}^{mem}$ and $I_n^{axial}$ denominate the radial and axial currents respectively \cite{Rall_1962}. The membrane currents are typically made up of external $I_{\text{ext}}$ and different internal ionic currents $I_{\text{ion}}$ such as sodium, potassium, and leak currents, yielding
\begin{equation}
    I^{mem} = I_{\text{ext}} + \sum_{\text{ion}} I_{\text{ion}}.\label{eq:membrane_currents}
\end{equation}
The axial currents are the result of the voltage dynamics of the connected compartments $m$,
\begin{equation}
    I^{n, axial} = \sum_m g_{n,m}(V_n - V_m), \label{eq:axial_currents}
\end{equation}
where the $V_m$ follows its own intrinsic dynamics subject to different internal and external currents $I_m^{mem}$. Following Hodgkin and Huxley \cite{hodgkin1952quantitative}, the ionic currents can be modeled as:
\begin{equation}
    I_{\text{ion}} = \overline{g}_{\text{ion}} \prod_{i} x_i^{k_i} \cdot (V - E_{\text{ion}}), \label{eq:ionic_current}
\end{equation}
with maximal conductance $\overline{g}_{\text{ion}}$, gating variables $x_i$ with gating exponent $k_i$, and reversal potential $E_{\text{ion}}$. A channel might have one or more gating variables, each of which follows its own dynamics, described by the first-order kinetics
\begin{equation}
    \frac{\text{d}x_i}{\text{d}t} = -\frac{x_i - x_{\infty}(V)}{\tau_x(V)}. \label{eq:gating_dynamics}
\end{equation}
$x_{\infty}(V)$ determines the steady-state at a given voltage, while $\tau_x(V)$ determines how fast the gate opens or closes.

\subsection{Neural ODEs}
Neural Ordinary Differential Equations (neural ODEs) parameterize continuous-time dynamical systems using neural networks, replacing discrete layer-wise transformations with a continuous evolution governed by an ordinary differential equation \cite{chen2018neural, kidger2021on}. Given a state variable $x(t) \in \mathbb{R}^d$, its dynamics are defined as
\begin{equation}
\frac{\text{d}x}{\text{d}t} = f_{\theta}(x(t), t),
\end{equation}
where $f_{\theta}$ is a neural network with parameters $\theta$. The solution defines a continuous flow map $\phi_t(x_0)$ such that $x(t) = \phi_t(x_0)$, obtained by numerically integrating the vector field from an initial condition $x_0$.
In practice, the evolution of the system is computed using standard ODE solvers (e.g., Runge--Kutta methods), which iteratively evaluate $f_{\theta}$. As a result, neural ODEs can naturally handle irregularly sampled data and variable time horizons, making them well-suited for modeling biophysical processes.

Model parameters are learned by minimizing a loss defined on observed trajectories. Gradients with respect to $\theta$ can be computed by differentiating through the ODE solver, either via direct backpropagation or the adjoint sensitivity method \cite{chen2018neural}, which solves a corresponding backward-in-time differential equation for the adjoint state. This enables memory-efficient training while maintaining compatibility with modern automatic differentiation frameworks.

\subsection{Hybrid Biophysical Models}\label{sec:hybrid_framework}
Universal differential equations \cite{rackauckas2020universal} provide a general framework for combining neural networks with mechanistic differential equations. We build on this idea with a domain-specific implementation for the biophysical setting. Within the conductance-based formalism of Sec.~\ref{sec:biophysical_models}, we replace or augment any current term in Eq.~\ref{eq:multicomp_formalism}---whether ionic or axial---by a neural ODE component, while preserving the remaining biophysical structure (Fig.~\ref{fig:graphical_abstract}). This yields a hybrid model of the form
\begin{equation}
    C \frac{\text{d}V}{\text{d}t} = \underbrace{I^{\text{known}}}_{\text{mechanistic}} + \underbrace{I_{\theta}^{\text{NODE}}}_{\text{learned}}, \label{eq:hybrid_model}
\end{equation}
which inherits the strong inductive biases of known ionic and axial currents $I^{\text{known}}$, while parameterizing the unknown currents, $I_{\theta}^{\text{NODE}}$, by a neural network with parameters $\theta$. This formulation allows us to describe any component of the model using either a biophysical or neural ODE formulation in a plug-and-play fashion. Depending on the application, the learned component can take different forms, which we describe below.

\paragraph{Learning gating dynamics of ionic currents.}
To model an unknown ionic current, we retain the standard conductance-based form (Eq.~\ref{eq:ionic_current}) but replace the assumed gating kinetics with a neural network. The learned current takes the form
\begin{align}
    I_{\theta}^{\text{NODE}} = \bar{g} \prod_{j} x_{\theta,j}^{k_j} \cdot (V - E), \qquad
    \frac{\text{d}x^{(j)}}{\text{d}t} = -\frac{x^{(j)} - x_{\infty,\theta}^{(j)}(V)}{\tau_{x,\theta}^{(j)}(V)}, \label{eq:node_channel}
\end{align}
where the maximal conductance $\bar{g}$, gating exponents $k_j$, and reversal potential $E$ can be learned or fixed to known values. A neural network outputs the voltage-dependent steady-state $x_{\infty,\theta}^{(j)}(V)$ and time constant $\tau_{x,\theta}^{(j)}(V)$ for each gating variable $x^{(j)}$. By parameterizing the neural ODE in terms of $x_\infty$ and $\tau_x$ rather than learning an unconstrained vector field, the model preserves the biophysical interpretation of gating variables: the recovered functions can be directly compared to experimental voltage-clamp measurements and to existing parametric channel models.

\paragraph{Learning effective axial currents.}
To capture the influence of dendritic morphology on somatic recordings without simulating the full multicompartment model, we introduce a learned axial current term, in the form of an augmented neural ODE \cite{Dupont_Doucet_Teh_2019}. Rather than recovering specific gating functions, we learn an effective current and its latent dynamics that summarize the aggregate dendritic contribution. We parameterize this as
\begin{equation}
    I_{\theta}^{\text{NODE}} = f_\theta(V, \mathbf{z}), \qquad \frac{\text{d}\mathbf{z}}{\text{d}t} = g_\theta(V, \mathbf{z}), \label{eq:node_axial}
\end{equation}
where $\mathbf{z} \in \mathbb{R}^d$ is a vector of latent variables whose dynamics are governed by a neural network $g_\theta$, and $f_\theta$ maps the current membrane voltage and latent state to an effective current. The latent variables $\mathbf{z}$ can be understood as a low-dimensional summary of the dendritic state that is not directly observed at the soma.

\paragraph{Training and Network Architecture.}
Training neural ODEs on stiff Hodgkin--Huxley-type dynamics requires careful design choices. We constrain the network outputs to biophysically valid ranges---sigmoid for $x_\infty \in (0,1)$ and exponential for $\tau_x > 0$ to cover the multiple orders of magnitudes that time constants can assume. This also regularizes the learned vector field and prevents pathological dynamics. Input voltages are normalized to $[-1, 1]$ and the derivatives are scaled to comparable magnitudes \cite{Kim_2021}. The training loss combines voltage MSE with a physics-informed derivative-matching term \cite{ghanem2024learning} that penalizes mismatch in the net current balance, providing gradient signal even when predicted and true spikes are temporally misaligned. To avoid non-spiking local minima, we use a multi-stage curriculum that progressively exposes the model to different dynamical characteristics of the recording, starting from the resting state, adding first a single action potential, then a pair of action potentials, and then the entire spike train. Neural network and biophysical parameters are optimized jointly using separate optimizers with different learning rates. Full architectural and training details, including hyperparameters and solver configuration, are provided in App.~\ref{app:architecture} and \ref{app:training}.

\section{Results}

\subsection{Neural Networks flexibly capture diverse ion channel kinetics}\label{sec:results_vclamp}
\begin{wrapfigure}{r}{0.5\textwidth}
  \begin{center}
    \includegraphics[width=0.48\textwidth]{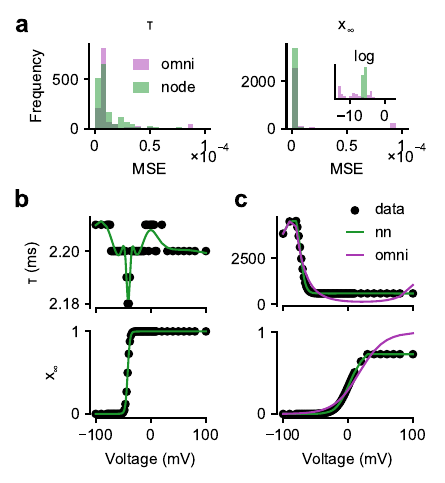}
  \end{center}
  \caption{Neural networks can represent gating dynamics derived from simulated voltage-clamp measurements. \textbf{a.} Prediction error for the steady state and time constant functions show no meaningful differences when parameterized by a neural network (nn) or the omnimodel (omni). \textbf{b.} Example fit for a complex gating function which could not be fit by the omnimodel. \textbf{c.} Example gating function that is better captured by the neural ODE.}
\label{fig:omni_fits}
\end{wrapfigure}
%
Before using neural ODEs to discover unknown gating dynamics from voltage recordings, we first verified that the neural network backbone can represent a wide range of ion channel kinetics. To this end, we fit a neural network to steady-state and time constant curves of simulated voltage clamp experiments for 2400 ion channel models from the IonChannelGenealogy database \cite{podlaski2017mapping}. The neural network architecture we used---a small multi-layer perceptron---is the same as the neural ODE backbone of our hybrid biophysical model in later experiments (experimental details in App.~\ref{app:omni_details}).
We compared against the omnimodel \cite{chintaluri2025omnimodel}, a parametric formulation that standardizes ion channel descriptions using a fixed functional form for $x_\infty(V)$ and $\tau_x(V)$.

Across these 2400 ion channels, the mean-squared error (MSE) for both $\tau(V)$ and $x_\infty(V)$ showed that the neural network achieved comparable performance to the omnimodel (Fig.~\ref{fig:omni_fits}a)---differences are so small they are practically negligible. Since the neural network does not assume a functional form, it was additionally able to fit about 180 gates that the omnimodel could not capture at all (Fig.~\ref{fig:omni_fits}b) and performed substantially better on channels with complex kinetics such as non-monotonic time constants or steady-state curves that deviate from canonical shapes (Fig.~\ref{fig:omni_fits}c), where the fixed parametric form of the omnimodel cannot fully capture the data. This establishes that the neural network architecture used in our hybrid model has sufficient expressiveness to represent the diversity of gating kinetics observed across a large variety of channel models and ion types. Hence, neural ODEs that parameterize steady-state and time constant curves (Eq.~\ref{eq:node_channel}) are expressive enough to imitate ion channels within biophysical simulations, which is a prerequisite for the mechanism discovery experiments that follow.
We also note that the neural network was easier to fit than the omnimodel, which required a much more elaborate fitting procedure (details in App.~\ref{app:omni_details}).

\subsection{Recovering gating dynamics from voltage recordings}\label{sec:results_recovery}
Having established that neural ODEs can represent diverse channel kinetics, we asked whether hybrid biophysical models can \emph{discover} unknown gating dynamics from voltage recordings alone. To this end, we simulated current-clamp data from a single compartment Hodgkin--Huxley model with sodium, potassium, and leak channels (details in App.~\ref{app:biophysical_models}). We then removed a channel from the model---either potassium or sodium---and replaced it with a synthetic neural ODE channel (Eqs.~\ref{eq:node_channel}), keeping the remaining channels fixed. The model received only a single voltage trace as a training signal; the gating variables of the unknown channel were never observed. The neural ODE had to learn the gating dynamics by inferring their underlying voltage-dependent functions ($x_\infty(V)$, $\tau_x(V)$) from the voltage signal alone.

\begin{figure}
    \centering
    \includegraphics[width=1.0\textwidth]{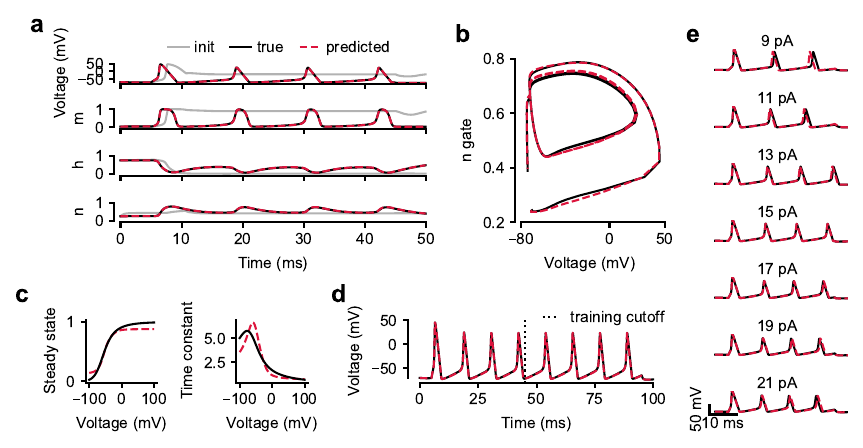}
    \caption{Recovering the dynamics of the K channel from a single current-clamp recording. \textbf{a.} The fitted voltage trajectories match the ground truth well. In addition, the dynamics of the n-gate were accurately recovered, despite the voltage trajectory being the only training signal. \textbf{b.} The predicted limit cycle follows the ground truth closely, indicating that the model has captured the underlying dynamics of the system. \textbf{c.} The inferred gating functions are close to the ground truth where the training data is dense, while deviating in regions with little training data. \textbf{d.} The inferred model is able to generalize to much longer stimuli. \textbf{e.} The model also generalizes well to unseen stimulus amplitudes far from the training stimulus of 15pA.}
    \label{fig:k_channel_recovery}
\end{figure}

Starting from a random initialization (gray traces), the model converged to accurate predictions of both the membrane voltage and all gating variables of the unknown channel (Fig.~\ref{fig:k_channel_recovery}a).
The recovered phase portraits closely matched the true limit cycle dynamics (Fig.~\ref{fig:k_channel_recovery}b), indicating that the model had learned the correct dynamical structure, not just a point-wise fit to the training trace.
The learned steady-state and time constant curves agreed well with the ground-truth functions (Fig.~\ref{fig:k_channel_recovery}c), demonstrating that the neural ODE parameterization in terms of $x_\infty$ and $\tau_x$ successfully recovered interpretable biophysical quantities without assuming their functional form.
While we also observed that the resulting fit differed from the ground-truth function at very low and high voltages, this can be explained by the fact that the dynamical system rarely takes on such voltages.

When simulated past the training cutoff, the voltage trace of the hybrid biophysical model remained accurate over a stimulus duration that was twice as long as what was seen during training (Fig.~\ref{fig:k_channel_recovery}d). In addition, the learned dynamics also generalized across stimulus amplitudes reproducing voltage traces for injected current amplitudes ranging from 9 to 21~pA, away from the 15~pA training stimulus (Fig.~\ref{fig:k_channel_recovery}e). We show that the same approach also accurately recovers the sodium channel, despite both the $m$ and $h$ gates needing to be inferred, with comparable performance and generalization abilities (Fig.~\ref{fig:na_channel_recovery}). Training details and loss curves can be found in App.~\ref{app:additional_results} and Fig.~\ref{fig:k_loss_curves}.
While an omnimodel could in principle be fit with the same setup, we were unable to achieve convergence with this setup, as the omnimodel was highly sensitive to initialization in the partially observed setting.

\begin{table}
\centering
\begin{tabular}{ccccc}
\hline
$\sigma_V$ & $\sigma_{\theta}$ & RMSE & min. RMSE & frac. converged \\
\hline
0.00 & 0.00 & \textbf{7.93 $\pm$ 17.05} & 0.63 & \textbf{0.85} \\
\hline
0.01 & 0.00 & 12.00 $\pm$ 19.81 & 0.51 & 0.75 \\
0.05 & 0.00 & 8.46 $\pm$ 16.39 & \textbf{0.38} & 0.80 \\
0.10 & 0.00 & 17.74 $\pm$ 19.77 & 0.65 & 0.40 \\
\hline
0.00 & 0.01 & 15.29 $\pm$ 22.62 & 0.57 & 0.70 \\
0.00 & 0.02 & 14.74 $\pm$ 21.45 & 0.46 & 0.70 \\
0.00 & 0.05 & 13.24 $\pm$ 17.33 & 1.03 & 0.50 \\
\hline
\end{tabular}
\caption{Robustness of the K channel recovery to observation noise ($\sigma_V$) and parameter misspecification ($\sigma_\theta$) across 20 runs per condition. RMSE is reported as mean $\pm$ std over all runs (in mV); min.\ RMSE is the best run. A run is considered converged if the final RMSE falls below 2\,mV. The model tolerates moderate noise ($\sigma_V \leq 0.05$) and parameter perturbations ($\sigma_\theta \leq 0.02$) with convergence rates of 70–80\%; convergence degrades to 40–50\% only under the most severe conditions tested.
}
\label{tab:k_channel_noise}
\end{table}

In practice, electrophysiological recordings are noisy and the biophysical parameters of the known model components may not be perfectly calibrated. We tested robustness to both factors by adding scaled Gaussian observation noise to the voltage traces ($\sigma_V \in \{0.01, 0.05, 0.1\}$) or perturbing the parameters of the known channels using scaled Gaussian noise ($\sigma_{\theta} \in \{0.01, 0.02, 0.05\}$). We considered a run converged if the final root mean squared error (RMSE) fell below 2 mV.

We found that the channel dynamics were recoverable even under observation noise and moderate parameter misspecification (Table~\ref{tab:k_channel_noise}). Noise up to $\sigma_V = 0.05$ had little effect: RMSE increased only marginally (8.46 vs.\ 7.93) and convergence remained at 80\%. At $\sigma_V = 0.1$, which substantially obscured the subthreshold dynamics (Fig.~\ref{fig:noise_examples}), convergence dropped to 40\% but successful runs still achieved low RMSE. Parameter misspecification up to $\sigma_\theta = 0.02$ was tolerated with 70\% convergence, and even at $\sigma_\theta = 0.05$---where the known channel parameters were perturbed enough to visibly alter spiking behavior (Fig.~\ref{fig:noise_examples}, bottom)---half of all runs still converged to the correct solution (Fig.~\ref{fig:noise_convergence_traces}). This indicated that the neural ODE in the hybrid biophysical model can absorb inaccuracies in the known model components, a key property when biophysical parameters are not known exactly. Importantly, these results were obtained with small neural networks that could be trained in about 10 minutes on a single CPU. In practice, running a small number of independent restarts---feasible given the ~10 min training time---substantially increases the probability of obtaining a converged solution. A comprehensive architecture sweep (Table~\ref{tab:arch_channel} in App.~\ref{app:additional_results}) showed that width $\geq 64$ was sufficient for 100\% convergence regardless of activation function or depth, making the approach practical and robust out of the box. Overall, these results show that our hybrid biophysical model can learn interpretable dynamics of ion channels from voltage measurements alone, and that the resulting dynamics generalize to longer time horizons, a wide range of stimulus amplitudes, and mis-specified biophysical models.

\subsection{Learning effective axial currents from morphologically detailed neurons}\label{sec:results_axial}
\begin{figure}[t]
    \centering
    \includegraphics[width=1.0\linewidth]{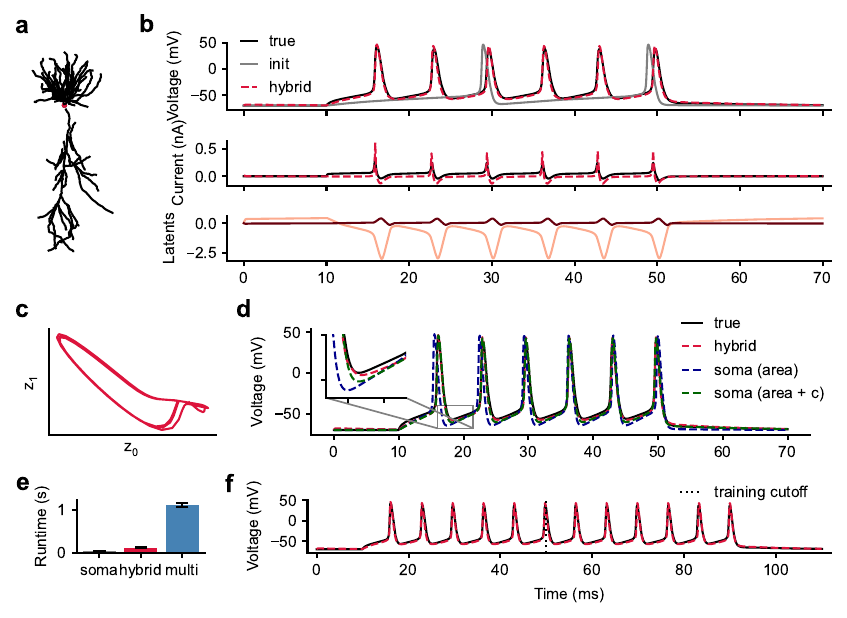}
    \caption{Emulating multicompartment dynamics using a single compartment model \textbf{a.} Reconstruction of a cortical neuron. Soma in red. \textbf{b.} First row: Despite initializing from a soma-only model (grey), the trained hybrid model matches data (red). Second row: The total axial current is recovered well by the model. Third row: The latents are modulated by the model in line with the steady state activity and action potentials. \textbf{c.} The activity of the latents resembles a limit cycle. \textbf{d.} Single compartment soma models, do not capture the shape of the action potentials as well as the hybrid biophysical model, even when membrane area and capacitance are trainable (blue, green). \textbf{e.} Simulation times for different models. Hybrid models are significantly faster than multicompartment models. \textbf{f.} The hybrid model generalizes well to stimuli twice as long as those seen during training.}
    \label{fig:multi_comp}
\end{figure}

Having shown that unknown ionic currents can be recovered in interpretable form, we now asked whether we can also account for the unmodeled aggregate influence of dendritic morphology. We aimed to learn effective axial currents to enable a single-compartment model to reproduce the somatic behavior of morphologically detailed neurons. To this end, we simulated the voltage response of a fully detailed CA1 pyramidal neuron from \citet{Pyapali_Sik_Penttonen_Buzsaki_Turner_1998} using realistic channels from \citet{Pospischil_2008}. We approximated the morphology using 154 compartments (Fig.~\ref{fig:multi_comp}a) and simulated it using \textsc{Jaxley} \citep{deistler2025jaxley}. We then built a hybrid biophysical model which consisted of a single-compartment somatic model augmented with a neural ODE, aiming to capture the axial currents (Eq.~\ref{eq:node_axial}) using 2 latent dimensions (details in App.~\ref{app:experimental_details}). We compared against a soma-only model with a learned membrane area, and a variant where both area and capacitance are adjusted.

The hybrid biophysical model closely matched the ground-truth voltage (Fig.~\ref{fig:multi_comp}b, red dashed), capturing spike timing, waveform shape, and subthreshold dynamics. This was achieved by inferring the axial currents of the soma compartment, which closely matched the true axial current (Fig.~\ref{fig:multi_comp}b, middle). In addition the model learned latent states (Fig.~\ref{fig:multi_comp}b, bottom) that exhibit structured, spike-correlated dynamics, suggesting the latent variables encode a meaningful summary of dendritic state. This is also evident from the limit cycle learned by the latents (Fig.~\ref{fig:multi_comp}c). This could not be achieved by a soma-only model with trainable morphological parameters, which showed inferior fits to the data (Fig.~\ref{fig:multi_comp}d). This confirmed that the dendritic inputs are necessary to reproduce somatic recordings from the full morphological model (Fig.~\ref{fig:multi_comp}a). In addition, the models are up to 10 times faster to simulate than their multicompartment counterparts, making it a practical surrogate for applications where simulation speed matters, such as large-scale network simulations or parameter sweeps. The overhead relative to a soma-only model is moderate, driven by the neural network evaluation within the ODE solver (Fig.~\ref{fig:multi_comp}e). Performance was evaluated across varying latent dimensions (1--8), network widths (32--512), and depths (2--3), which revealed that the axial currents can be recovered reliably across different setups (Tab.~\ref{tab:arch_multicomp}). Finally, we tested the ability of the hybrid biophysical model to extrapolate beyond the training window. We found that it accurately predicted somatic voltage for a stimulus duration twice as long as what was seen during training (Fig.~\ref{fig:multi_comp}f), demonstrating that it captured meaningful dendritic contributions.

\section{Discussion}\label{sec:discussion}
We introduced hybrid biophysical models, a plug-and-play framework that embeds neural ODEs into conductance-based neuron models to recover unknown ionic and axial current dynamics from voltage recordings. Any component of a biophysical model can be selectively replaced with a neural ODE while the rest remains mechanistic, making the approach compatible with existing models and simulators. We demonstrated the power of these models across three settings: fitting channel kinetics of more than 2000 channels, recovering gating dynamics from single current-clamp recordings under noise and parameter misspecification, and reducing a 154-compartment model to a single-compartment surrogate with learned axial currents at up to an order of magnitude lower computational cost.

Learning Hodgkin--Huxley-like dynamics with neural ODEs is challenging due to the stiffness of the system, which neural ODEs are known to struggle with \cite{Kim_2021, matzakos2026comparing, Ioannou_2026}. We addressed this through four design choices: an $x_\infty$/$\tau_x$ parameterization that constrains the learned vector field to biophysically valid ranges and avoids pathological dynamics; input and equation scaling \cite{huang2025training, Kim_2021}; a hybrid loss with a derivative-matching term that provides gradient signal even when spikes are misaligned \cite{raissi2019physics, sholokhov2023pinode}; and a curriculum that exposes the model to progressively longer traces. Together these choices enabled reliable convergence for sufficiently large networks, with models training in 10 minutes on a single CPU. While training hybrid biophysical models requires that the biophysical simulator is fully differentiable, the final learned channel models can be plugged into any standard conductance-based simulator \cite{deistler2025jaxley, hines1997neuron} either directly or by first tabulating discovered gating dynamics $x_\infty(V)$ and $\tau_x(V)$ on a voltage grid, even without the neural network. If a compact analytical description is preferred, the learned curves can be further distilled into closed-form expressions via symbolic regression \cite{cranmer2023interpretable, rackauckas2020universal} or by fitting a parametric model \cite{chintaluri2025omnimodel}. This separation of discovery from description is a key advantage---the neural ODE handles the hard problem of finding the right dynamics without committing to a functional form, and Occam's razor can be applied afterwards to select the simplest model that captures them.

Despite these strengths, recovering gating dynamics from voltage alone is inherently underconstrained. For channels with multiple gates (e.g., Na with $m^3 h$), multiplicative interactions allow different combinations of $x_\infty$ and $\tau_x$ to produce similar currents without additional data or external regularization. The conductance-based formalism of our hybrid biophysical models mitigates this compared to an unconstrained neural ODE (Fig.~\ref{fig:na_channel_recovery}): the recovered functions matched ground truth well, deviating only at tails where training data was sparse. When biophysical parameters must be co-optimized the system will become further underdetermined \cite{Casolo_2025}. The situation is different for the axial current formulation: here, the latent variables are not intended to recover specific biophysical quantities but serve as a surrogate for the aggregate dendritic contribution. Hence they do not reflect a mechanistic account of dendritic activity. Working exclusively on simulated data allowed us to validate against known ground truth---a necessary first step before applying the framework to real recordings. This will require addressing additional challenges, such as unmodeled noise sources and unknown channel composition. In such settings, quantifying uncertainty over the learned gating functions will become essential, which our approach can be naturally extended to support, for example with Bayesian neural ODEs \cite{Dandekar_2022}, ensembles \cite{lakshminarayanan2017simple}, or post-hoc calibration methods \cite{ott2023uncertainty}. Looking further ahead, differentiable simulators \cite{deistler2025jaxley} open the door to applying our framework beyond single compartments: neural ODE components could be inserted into select compartments of a morphologically detailed model to learn the dynamics of specific regions while keeping the rest mechanistic. Finally, training a shared model across many channels or neurons in an amortized fashion could substantially reduce fitting cost and enable in-context learning, adapting to new cells or channel types without retraining from scratch.

Our framework targets a sweet spot between fully mechanistic and fully data-driven models: it preserves biophysical structure where known and learns what is missing, with recovered dynamics expressed in interpretable biophysical terms. Its plug-and-play nature allows selectively learning any membrane or axial current and its underlying dynamics, while leaving the remaining model components mechanistic. Crucially, it is precisely these mechanistic constraints that guide the neural ODE toward physically meaningful representations---as evidenced by the ability to generalize to unseen stimulus regimes and extrapolate well beyond the training window. We have shown this to be robust across noise, misspecification, and architecture settings, establishing hybrid biophysical models as a practical tool for mechanism discovery in neuroscience.

\section*{Acknowledgments}
We thank colleagues, especially Cornelius Schröder and Daniel Gedon for helpful discussions at early stages of this work. We thank Chaitanya Chintaluri for his help in comparing against the omni model. We thank the International Max Planck Research School for Intelligent Systems (IMPRS-IS) for supporting JB.
This work was supported by the German Research Foundation (DFG) through Germany’s Excellence Strategy (EXC-Number 2064/1, PN 390727645) and SFB1233 (PN 276693517), and the European Union (ERC, DeepCoMechTome, 101089288). Views and opinions expressed are however those of the author(s) only and do not necessarily reflect those of the European Union or the European Research Council. Neither the European Union nor the granting authority can be held responsible for them.

\bibliographystyle{unsrtnat}
\bibliography{references}

\clearpage
\appendix

\setcounter{section}{0}
\renewcommand{\thesection}{\Alph{section}}

\renewcommand{\thefigure}{\Alph{section}-\arabic{figure}}
\setcounter{figure}{0}
\renewcommand{\thetable}{\Alph{section}-\arabic{table}}
\setcounter{table}{0}
\renewcommand{\theequation}{\Alph{section}-\arabic{equation}}
\setcounter{equation}{0} 

\section{Experimental Details}\label{app:experimental_details}
In the following we detail the network architecture, training procedure, and biophysical models used in all experiments. All experiments were implemented in JAX \cite{jax2018github} using Equinox \cite{kidger2021equinox} for neural network modules, Diffrax \cite{kidger2021on} for differentiable ODE solvers, and \textsc{Jaxley} \cite{deistler2025jaxley} to generate some of the training data. All simulations and training runs were performed on CPUs (Intel Xeon Gold 6342 cluster nodes or a laptop with an 11th Gen Intel Core i7-1165G7 and 16\,GB RAM). 

Code and instructions for how to run the experiments can be found at \url{https://github.com/berenslab/hybrid_biophysical_models_preprint}.

\subsection{Network Architecture}\label{app:architecture}
Each gating variable $x_i$ is assigned a dedicated MLP whose sole input is the membrane voltage. This enforces conditional independence between gates and avoids cross-gate interference during optimization, which is particularly important when recovering multiple gates simultaneously (e.g., $m$ and $h$ for Na; Fig.~\ref{fig:na_channel_recovery}).
 
Before being passed to the network, voltage is linearly scaled to $[-1, 1]$: $u = 2\,(V - V_{\min}) / (V_{\max} - V_{\min}) - 1$ with $V_{\min} = -80\,\text{mV}$, $V_{\max} = 60\,\text{mV}$, which improves training stability. For the gating recovery experiments (Sec.~\ref{sec:results_vclamp} and \ref{sec:results_recovery}), the network predicts two outputs per gate: the steady state $x_{\infty,i} = \sigma(a_i) \in (0, 1)$ via a sigmoid, and the time constant $\tau_i = \exp(b_i) > 0$ via an exponential, where $(a_i, b_i)$ are the raw network outputs. This ensures the steady state is constrained to $[0, 1]$ and allows the time constant to vary across multiple orders of magnitude, as is common for many gates. To avoid numerical issues with products of small gate values raised to potentially large exponents, the ionic current is computed in log-space: $I = g_x \exp\!\bigl(\sum_i p_i \ln x_i\bigr)(V - E_x)$, where $g_x$, $E_x$, and the exponents $p_i$ can be learnable scalar parameters but were initialized the known values of the channel that was being modeled.
 
For the axial current experiments (Sec.~\ref{sec:results_axial}), the input voltage is scaled by $1/100$ and the output predicting the current is scaled by a factor of 10. The latent variables are left unscaled.
 
We use \texttt{tanh} activations throughout as the default, due to their stability and being fast to train (Tab.~\ref{tab:arch_channel}). Other activations such as \texttt{softplus} and \texttt{mish} were also evaluated (see Tables~\ref{tab:arch_channel} and \ref{tab:arch_multicomp}) and were sometimes better at fitting the data, particularly in stiff scenarios. Layer normalization is applied after each hidden layer to stabilize training across the wide range of time constants encountered.

All computations are performed in 64-bit floating point, which is required for numerically stable integration of the stiff Hodgkin--Huxley dynamics.

The results for the K channel recovery (Fig.~\ref{fig:k_channel_recovery}) were obtained using an MLP with width 32, depth 2, and \texttt{tanh} activation. For the Na channel recovery (Fig.~\ref{fig:na_channel_recovery}), we used width 128, depth 3, and \texttt{silu} activation, reflecting the increased complexity of simultaneously recovering two interacting gates. To emulate the axial currents in Fig.~\ref{fig:multi_comp}, we used width 512, depth 2, with 2 latent dimensions.

\subsection{Training}\label{app:training}
\paragraph{Gate fitting (voltage-clamp data).} For Sec.~\ref{sec:results_vclamp}, the network is trained to minimize the mean squared error on the gate steady-state $x_\infty(V)$ and time constant $\tau(V)$, with each output independently scaled by $\max(\mathrm{std}(\cdot), 1)$ to prevent one output from dominating the loss. Network weights are initialized from a normal distribution. A two-stage learning rate schedule is used: (1)~warmup from $10^{-6}$ to $10^{-3}$ over 1\,000 steps, held constant for 10\,000 total steps; (2)~warmup followed by cosine decay to zero over 10\,000 steps. AdamW \cite{loshchilov2018decoupled} is used with weight decay $10^{-4}$. The setup was not specifically optimized for this task.
 
\paragraph{Voltage fitting (current-clamp data).} For Sec.~\ref{sec:results_recovery} and \ref{sec:results_axial}, the loss combines a voltage-matching term with a physics-informed derivative-matching term:
\begin{equation}
    \mathcal{L} = \underbrace{\frac{1}{T}\sum_t \bigl(V_\theta(t) - V(t)\bigr)^2}_{\text{voltage MSE}} + \;\lambda_1 \underbrace{\frac{1}{T}\sum_t \Bigl(\tfrac{\text{d}V_\theta}{\text{d}t}\big|_t - \tfrac{\text{d}V}{\text{d}t}\big|_t\Bigr)^2}_{\text{derivative MSE}},
    \label{eq:loss}
\end{equation}
where time derivatives are estimated via finite differences and $\lambda_1 = 10^{-2}$ by default. The derivative term penalizes mismatch in the net ionic current balance ($C\,dV/dt = I_\mathrm{ext} + \sum I_\mathrm{ion}$), providing an additional supervisory signal beyond point-wise voltage agreement. This is particularly important early in training when predicted and true spikes may be temporally misaligned.
 
\paragraph{Initialization and steady-state regularization.} Before training begins, the latent gate variables are initialized to their predicted steady-state values at the resting membrane potential: the network is evaluated at $V_0$ to obtain $x_{\infty,\theta}(V_0)$, which is used as the initial condition $x_i(0) = x_{\infty,i}(V_0)$. During the first curriculum stage, a \textit{force-state} regularizer additionally penalizes deviations of the predicted trajectory from the target resting state (both voltage and latents), ensuring that the latent variables do not drift from their steady-state values before stimulus onset.
 
\paragraph{Optimizer.} Neural network parameters are optimized with AdamW; scalar (biophysical) parameters (such as the learned stimulus amplitude in Sec.~\ref{sec:results_axial}) are optimized with Adam. Gradient updates from the two parameter groups are computed simultaneously via a multi-transform optimizer with separate learning rates. A global gradient norm clip of $0.1$ is applied. Weight decay is set to $10^{-4}$ for gate fitting and $10^{-5}$ for voltage fitting. Voltage fitting uses a cosine one-cycle schedule per training stage (peak learning rates $10^{-1} \to 10^{-4} \to 10^{-5}$ across stages).
 
\paragraph{Multi-stage curriculum.} Voltage-trajectory training proceeds in stages that progressively expose longer portions of the recording:
\begin{enumerate}
    \item \textit{Steady-state stage}: the force-state regularizer is active and the truncated window ends just before stimulus onset, training the model to reproduce stable resting dynamics.
    \item \textit{First action potential stage}: the window extends to cover the first action potential, ensuring the model is driven to spike and preventing the optimizer from settling on a non-spiking local minimum.
    \item \textit{Full-trace stage}: the model is trained on the complete recording to reproduce the full spiking pattern and inter-spike dynamics.
\end{enumerate}
Multicompartment experiments (Sec.~\ref{sec:results_axial}) use a finer curriculum with up to six stages that additionally include an initial current amplitude scaling phase and intermediate stages that progressively add individual action potentials (see Fig.~\ref{fig:multicompartment_losses} for the corresponding loss curves).
 
\paragraph{Bounded scalar parameters.} Scalar (biophysical) parameters are constrained to physically plausible ranges using a logistic transform that maps the unconstrained optimization variable to a specified $[\text{lower}, \text{upper}]$ interval. This avoids constrained optimization while keeping parameters in bounds throughout training.

\paragraph{ODE solver.} Trajectories are integrated with Tsit5 (explicit 4th/5th-order Runge--Kutta) with adaptive step-size control via a PID controller ($\mathrm{rtol} = 10^{-6}$, $\mathrm{atol} = 10^{-8}$). Gradients are computed through time and with respect to both the biophysical and neural network parameters via a checkpointed adjoint sensitivity method \cite{chen2018neural}, which checkpoints intermediate states to limit memory usage during backpropagation through the ODE solver.
 
\subsection{Biophysical Models}\label{app:biophysical_models}
 
\paragraph{Hodgkin--Huxley model (Sec.~\ref{sec:results_recovery}).} We use a standard single-compartment Hodgkin--Huxley model \cite{hodgkin1952quantitative} with sodium, potassium, and leak currents:
\begin{equation}
    C\frac{\text{d}V}{\text{d}t} = I_{\text{ext}} - \bar{g}_{\text{Na}}\,m^3 h\,(V - E_{\text{Na}}) - \bar{g}_{\text{K}}\,n^4\,(V - E_{\text{K}}) - g_{\text{leak}}\,(V - E_{\text{leak}})
    \label{eq:hh_full}
\end{equation}
with standard rate functions $\alpha_z(V)$ and $\beta_z(V)$ for $z \in \{m, h, n\}$ \cite{hodgkin1952quantitative}. Parameters: $\bar{g}_{\text{Na}} = 120$, $\bar{g}_{\text{K}} = 36$, $g_{\text{leak}} = 0.3\,\text{mS/cm}^2$; $E_{\text{Na}} = 50$, $E_{\text{K}} = -75$, $E_{\text{leak}} = -70\,\text{mV}$; $C = 1\,\mu\text{F/cm}^2$. The optimization target was generated by simulating the voltage response to a step current of $I_{\text{ext}} = 15\,pA$, spread over $A_{\text{soma}}=62.8\,\mu m^2$, from $t = 5\,\text{ms}$ to $t = 45\,\text{ms}$, with total duration $50\,\text{ms}$ at $\Delta t = 0.1\,\text{ms}$, initialized from $V_0 = -70\,\text{mV}$ with all gating variables at their steady-state values.
 
\paragraph{Multicompartment model (Sec.~\ref{sec:results_axial}).} We consider a CA1 pyramidal neuron morphology \cite{Pyapali_Sik_Penttonen_Buzsaki_Turner_1998}, discretized into 154 compartments (one per branch) with axial resistivity $R_a = 100\,\Omega\cdot\text{cm}$. Somatic membrane dynamics follow the Regular Spiking (RS) model of \citet{Pospischil_2008}, which includes sodium, potassium, leak, and a slow non-inactivating K$^+$ M-current:
\begin{equation}
    C\frac{dV}{dt} = I_{\text{ext}} - \bar{g}_{\text{Na}}\,m^3 h\,(V - E_{\text{Na}}) - \bar{g}_{\text{K}}\,n^4\,(V - E_{\text{K}}) - g_{\text{leak}}\,(V - E_{\text{leak}}) - \bar{g}_M\,p\,(V - E_{\text{K}})
    \label{eq:pospischil}
\end{equation}
with rate constants $\alpha_z(V)$ and $\beta_z(V)$ for $z \in \{m, h, n, p\}$ as specified in \citet{Pospischil_2008}. Parameters: $\bar{g}_{\text{Na}} = 50$, $\bar{g}_{\text{K}} = 5$, $g_{\text{leak}} = 0.1$, $\bar{g}_M = 0.004\,\text{mS/cm}^2$; $E_{\text{Na}} = 50$, $E_{\text{K}} = -90$, $E_{\text{leak}} = -70\,\text{mV}$; $C = 1\,\mu\text{F/cm}^2$; M-current time constant $\tau_{\max} = 4000\,\text{ms}$; voltage threshold $V_T = -60\,\text{mV}$. Target traces were generated by injecting a somatic step current of $0.65\,\text{nA}$ ($\approx 26\,\mu\text{A/cm}^2$ at the soma) from $t = 10\,\text{ms}$ to $t = 50\,\text{ms}$, with total duration $70\,\text{ms}$.

\subsection{Fitting Ion Channel Kinetics from Voltage-Clamp Data}\label{app:omni_details}

The data for the experiments in Sec.~\ref{sec:results_vclamp} were obtained from the IonChannelGenealogy (ICG) database \cite{podlaski2017mapping}, specifically from the curated data files at \url{https://github.com/icgenealogy/icg-channels/tree/8179f57/icg-pickles}. Each file contains the numerically extracted steady-state activation $x_\infty(V)$ and time constant $\tau_x(V)$ curves for a single gating variable, obtained by running the original NEURON \cite{hines1997neuron} channel model files (.mod) under standard voltage-clamp protocols at 6.3\textdegree C. The database contains 3,524 voltage-gated channel models spanning five ion classes: potassium (K, 1,455), sodium (Na, 923), calcium (Ca, 665), I$_h$ (246), and calcium-activated potassium (KCa, 235). Of these, we selected a subset of 2400 channels for which the data was consistently formatted and did not have any errors.

\paragraph{Omnimodel baseline.} We compare against the omnimodel \cite{chintaluri2025omnimodel}, which follows \citet{destexhe2000which} and parameterizes the steady-state as a modified sigmoid with four parameters:
\begin{equation}
    x_\infty(V) = \frac{c}{1 + \exp(-aV + b)} + d,
\end{equation}
where $a$ controls the slope, $b$ sets the half-activation voltage, $c$ is a multiplicative scaling, and $d$ an offset. The time constant is parameterized as a fourth-order thermodynamic formulation with ten parameters $(v_h, A, b_1, c_1, d_1, b_2, c_2, d_2, e_1, e_2)$:
\begin{equation}
    \tau_x(V) = \frac{A}{B + C},
\end{equation}
where
\begin{align}
    B &= e^{-[b_1(V - v_h) + c_1(V - v_h)^2 + d_1(V - v_h)^3 + e_1(V - v_h)^4]}, \\
    C &= e^{[b_2(V - v_h) + c_2(V - v_h)^2 + d_2(V - v_h)^3 + e_2(V - v_h)^4]}.
\end{align}
To assess whether the full fourth-order model is needed, \citet{chintaluri2025omnimodel} incrementally increase the polynomial order by setting higher-order coefficients to zero, fitting up to five models of increasing complexity per gate. For each gating variable, $x_\infty$ and $\tau_x$ are fit independently via non-linear least squares, and the best $\tau_x$ fit is retained.

\paragraph{Neural ODE fitting and comparison.} Our neural ODE fits $x_\infty$ and $\tau_x$ jointly using a single MLP per gate (width 32, depth 2, \texttt{tanh}), trained via gradient descent with combined MSE on both curves. We use one architecture for all channels with no channel-specific tuning. In Fig.~\ref{fig:omni_fits}, we compare against the best of the five omnimodel fits per channel. Despite this favorable setup for the baseline, the neural ODE achieves comparable MSE across channels and can additionally fit cases that none of the five parametric forms could capture.

\section{Additional Results}\label{app:additional_results}
 
\subsection{Recovering Gating Dynamics from Voltage Recordings}
 
\begin{figure}[h]
    \centering
    \includegraphics[width=1.0\textwidth]{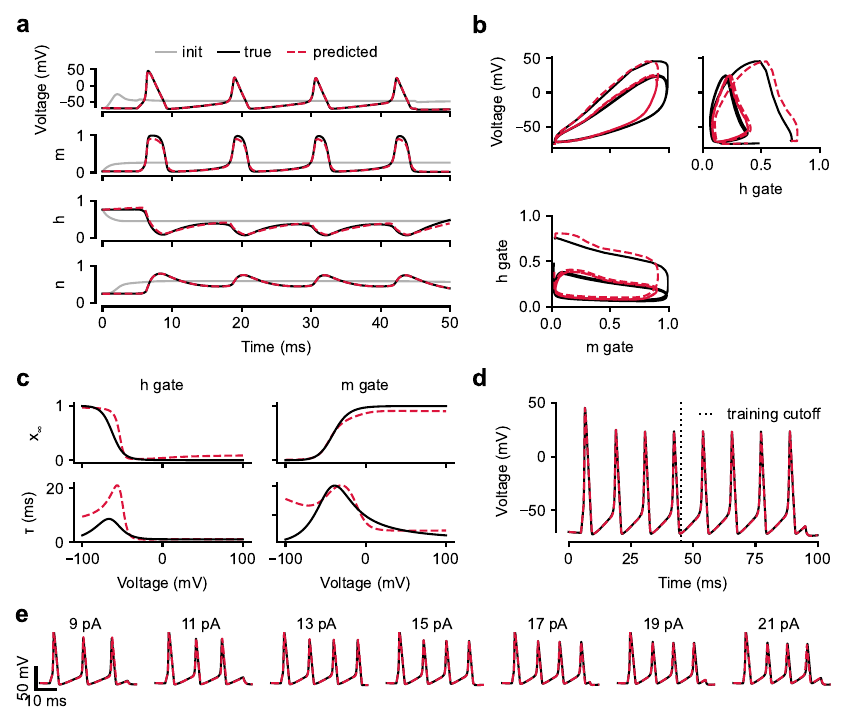}
    \caption{Recovering the dynamics of the Na channel from a single current-clamp recording. \textbf{a.}~The fitted voltage trajectories closely match the ground truth, and the dynamics of both the $m$- and $h$-gates are recovered well despite only the voltage being observed. \textbf{b.}~The predicted limit cycle captures the dynamics well too, despite the multiplicative interaction between gates. \textbf{c.}~The inferred gating functions are close to the ground truth, with only the time constants deviating significantly in the hyperpolarized regimes, for which no training data was present. \textbf{d.}~The model generalizes to stimuli twice as long as those seen during training. \textbf{e.}~The model generalizes well to unseen stimulus amplitudes far from the training stimulus of 15\,pA.}
    \label{fig:na_channel_recovery}
\end{figure}

\paragraph{Sodium channel recovery.} In the main text we show recovery of the potassium channel (Fig.~\ref{fig:k_channel_recovery}). Here we present the analogous experiment for the sodium channel, where both the $m$ and $h$ gating variables must be recovered simultaneously. The fitted voltage trajectories closely match the ground truth, and both gate dynamics are recovered well (Fig.~\ref{fig:na_channel_recovery}a). The predicted limit cycle does not perfectly follow the ground truth due to the multiplicative interaction between $m$ and $h$, but the general dynamics are preserved (Fig.~\ref{fig:na_channel_recovery}b). The inferred gating functions are close to the ground truth, with the exception of the $m$-gate time constant which shows some deviation (Fig.~\ref{fig:na_channel_recovery}c). Despite these imperfections, the model generalizes to stimuli twice as long as the training window (Fig.~\ref{fig:na_channel_recovery}d) and to unseen stimulus amplitudes far from the 15\,pA training stimulus (Fig.~\ref{fig:na_channel_recovery}e).
 
\begin{figure}[h]
    \centering
    \includegraphics[width=1.0\linewidth]{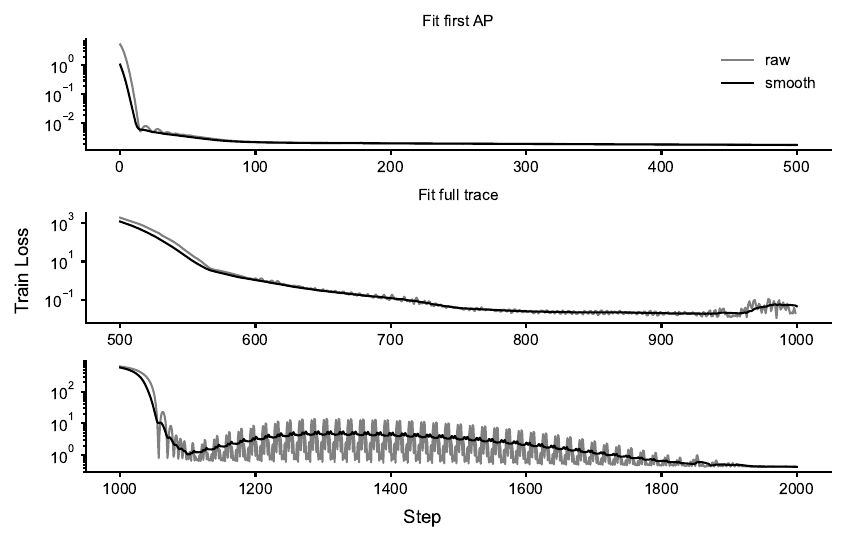}
    \caption{Training loss curves for the K channel recovery experiment (Sec.~\ref{sec:results_recovery}). The model first learns to fit the initial action potential (top), then the full trace including all remaining spikes (bottom). The highly non-smooth loss landscape reflects the difficulty of aligning predicted and true action potentials; sudden drops correspond to the predicted spikes snapping into alignment with the data. The general bell curve shape can be attributed to the cosine one cycle learning rate scheduling.}
    \label{fig:k_loss_curves}
\end{figure}

\paragraph{Training dynamics.} The loss curves for the K channel recovery (Fig.~\ref{fig:k_loss_curves}) illustrate the challenging optimization landscape of this problem. The model first learns to fit the initial action potential, after which the remaining spikes are learned in sequence. The loss curve is highly non-smooth, with sudden drops corresponding to moments when predicted action potentials abruptly align with the training data. This behavior motivates the curriculum training strategy described in App.~\ref{app:training}.
 
\begin{figure}[h]
    \centering
    \includegraphics[width=1.0\linewidth]{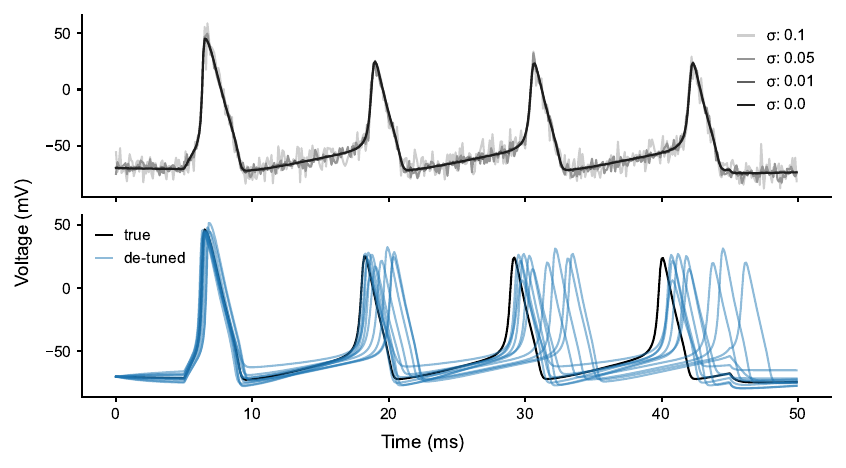}
    \caption{Effect of noise and parameter misspecification on the voltage traces. \textbf{Top:}~Different levels of observation noise $\sigma_V$ added to the voltage trace. \textbf{Bottom:}~The effect of mis-tuned biophysical parameters ($\sigma_{\theta_0} = 0.1$) on the model output. Each trace corresponds to a different random perturbation of the known channel parameters; the neural ODE must compensate for this mismatch since the training data was generated from the unperturbed (``true'') model.}
    \label{fig:noise_examples}
\end{figure}

\paragraph{Noise and parameter misspecification.} Figure~\ref{fig:noise_examples} illustrates the magnitude of the noise and parameter perturbations used in the robustness experiments of Tab.~\ref{tab:k_channel_noise}. The top panel shows voltage traces corrupted with different levels of observation noise $\sigma_V$; even at $\sigma_V = 0.05$ the spike waveforms remain clearly identifiable, while at $\sigma_V = 0.1$ the noise substantially obscures the subthreshold dynamics. The bottom panel shows the effect of mis-tuned biophysical parameters ($\sigma_{\theta_0} = 0.1$) on the ground-truth model: different parameter perturbations lead to dramatically different voltage traces, with spikes shifting in timing, amplitude, and frequency. The neural ODE must compensate for this mismatch, since the training data was generated from the unperturbed model.

\begin{figure}[h]
    \centering
    \includegraphics[width=1.0\linewidth]{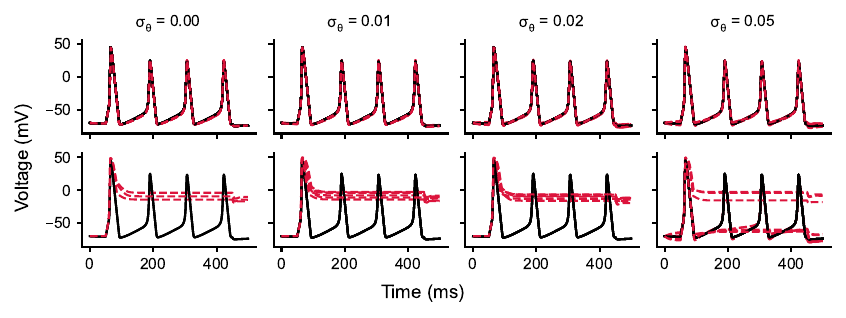}
    \caption{Fitted voltage traces for different levels of parameter misspecification $\sigma_\theta$. \textbf{Top row:}~Converged runs closely match the ground truth (black) across all noise levels, showing that the neural ODE can absorb parameter inaccuracies in the known channels. \textbf{Bottom row:}~Failed runs settle on non-spiking or subthreshold solutions. The fraction of failed runs increases with $\sigma_\theta$ (from 16\% at $\sigma_\theta = 0$ to 50\% at $\sigma_\theta = 0.05$; see Table~\ref{tab:k_channel_noise}).}
    \label{fig:noise_convergence_traces}
\end{figure}

Figure~\ref{fig:noise_convergence_traces} shows representative fitted voltage traces across different levels of parameter misspecification $\sigma_\theta$. At $\sigma_\theta = 0$ (no misspecification), all converged runs closely track the ground truth across the full trace. As $\sigma_\theta$ increases, the converged runs (top row) continue to produce accurate fits, demonstrating that the neural ODE successfully compensates for the mis-tuned known channels. However, the fraction of runs that fail to converge grows with $\sigma_\theta$: failed runs (bottom row) typically settle on non-spiking or low-amplitude solutions, particularly at $\sigma_\theta = 0.05$ where only 50\% of runs converged. This illustrates that parameter misspecification primarily affects convergence reliability rather than the quality of successful fits.

\paragraph{Architecture sweep.} Table~\ref{tab:arch_channel} reports the results of a systematic sweep over network width, depth, and activation function for the K channel recovery task. We define convergence as achieving a final RMSE below 2\,mV, at which point all solutions practically have the correct number of spikes at the right place and only small misalignment errors remain. Networks with width $\geq 64$ achieve 100\% convergence across all activation functions, while smaller networks show highly variable performance. Among compact architectures, \texttt{tanh} with width 32 and depth 3 is a notable outlier that achieves full convergence. Training time scales roughly linearly with network size, ranging from $\sim$5 minutes for the smallest networks to $\sim$20 minutes for width 128.

\begin{table}[h]
\centering
\begin{tabular}{ccccccc}
\hline
width & depth & activation & RMSE & min. RMSE & frac. converged & runtime (min) \\
\hline
8 & 2 & mish & 20.63 $\pm$ 4.49 & 5.06 & 0.00 & 5.07 $\pm$ 0.58 \\
8 & 2 & silu & 19.97 $\pm$ 2.40 & 13.14 & 0.00 & 5.18 $\pm$ 0.44 \\
8 & 2 & softplus & 19.05 $\pm$ 4.91 & 6.21 & 0.00 & 5.32 $\pm$ 0.58 \\
8 & 2 & tanh & 17.44 $\pm$ 6.32 & 4.60 & 0.09 & 5.21 $\pm$ 0.33 \\
8 & 3 & mish & 22.23 $\pm$ 2.15 & 19.64 & 0.00 & 5.49 $\pm$ 0.56 \\
8 & 3 & silu & 19.30 $\pm$ 4.70 & 4.76 & 0.05 & 5.47 $\pm$ 0.63 \\
8 & 3 & softplus & 18.85 $\pm$ 5.53 & 4.70 & 0.04 & 5.50 $\pm$ 0.63 \\
8 & 3 & tanh & 21.17 $\pm$ 4.75 & 12.02 & 0.00 & 5.66 $\pm$ 0.61 \\
16 & 2 & mish & 28.31 $\pm$ 17.84 & 3.34 & 0.10 & 4.86 $\pm$ 0.46 \\
16 & 2 & silu & 30.41 $\pm$ 19.42 & 4.71 & 0.06 & 4.82 $\pm$ 0.52 \\
16 & 2 & softplus & 34.38 $\pm$ 18.33 & 12.18 & 0.00 & 4.77 $\pm$ 0.41 \\
16 & 2 & tanh & 39.10 $\pm$ 17.70 & 13.19 & 0.00 & \textbf{4.70 $\pm$ 0.50} \\
16 & 3 & mish & 20.49 $\pm$ 19.79 & 2.14 & 0.32 & 5.62 $\pm$ 0.71 \\
16 & 3 & silu & 13.85 $\pm$ 10.80 & 2.40 & 0.37 & 5.95 $\pm$ 0.68 \\
16 & 3 & softplus & 25.57 $\pm$ 23.36 & 2.40 & 0.32 & 5.55 $\pm$ 0.75 \\
16 & 3 & tanh & 19.14 $\pm$ 16.06 & 1.52 & 0.26 & 5.47 $\pm$ 0.80 \\
32 & 2 & mish & 18.41 $\pm$ 25.60 & 0.39 & 0.62 & 5.76 $\pm$ 0.92 \\
32 & 2 & silu & 11.60 $\pm$ 20.75 & 0.48 & 0.76 & 6.10 $\pm$ 0.79 \\
32 & 2 & softplus & 8.84 $\pm$ 19.66 & \textbf{0.32} & 0.82 & 6.12 $\pm$ 0.63 \\
32 & 2 & tanh & 4.71 $\pm$ 14.01 & 0.34 & 0.91 & 6.42 $\pm$ 0.58 \\
32 & 3 & mish & 17.85 $\pm$ 24.99 & 0.36 & 0.65 & 6.87 $\pm$ 1.03 \\
32 & 3 & silu & 12.22 $\pm$ 22.46 & 0.55 & 0.75 & 6.89 $\pm$ 0.95 \\
32 & 3 & softplus & 14.81 $\pm$ 24.86 & 0.60 & 0.72 & 6.69 $\pm$ 0.94 \\
32 & 3 & tanh & 1.03 $\pm$ 0.37 & 0.53 & \textbf{1.00} & 7.48 $\pm$ 0.49 \\
64 & 2 & mish & 1.20 $\pm$ 0.27 & 0.84 & \textbf{1.00} & 7.82 $\pm$ 0.64 \\
64 & 2 & silu & \textbf{1.01 $\pm$ 0.38} & 0.52 & \textbf{1.00} & 7.56 $\pm$ 0.62 \\
64 & 2 & softplus & 1.16 $\pm$ 0.43 & 0.55 & \textbf{1.00} & 7.73 $\pm$ 0.65 \\
64 & 2 & tanh & 1.04 $\pm$ 0.34 & 0.50 & \textbf{1.00} & 7.89 $\pm$ 0.64 \\
64 & 3 & mish & 1.40 $\pm$ 0.26 & 1.03 & \textbf{1.00} & 9.47 $\pm$ 0.96 \\
64 & 3 & silu & 1.20 $\pm$ 0.22 & 0.90 & \textbf{1.00} & 9.25 $\pm$ 1.02 \\
64 & 3 & softplus & 1.24 $\pm$ 0.34 & 0.67 & \textbf{1.00} & 9.52 $\pm$ 0.96 \\
64 & 3 & tanh & 1.31 $\pm$ 0.36 & 0.60 & \textbf{1.00} & 9.85 $\pm$ 1.02 \\
128 & 2 & mish & 1.57 $\pm$ 0.59 & 0.69 & \textbf{1.00} & 12.53 $\pm$ 2.56 \\
128 & 2 & silu & 1.49 $\pm$ 0.34 & 0.72 & \textbf{1.00} & 13.58 $\pm$ 1.84 \\
128 & 2 & softplus & 1.49 $\pm$ 0.35 & 0.86 & \textbf{1.00} & 13.07 $\pm$ 2.28 \\
128 & 2 & tanh & 1.58 $\pm$ 0.44 & 0.97 & \textbf{1.00} & 12.41 $\pm$ 2.45 \\
128 & 3 & mish & 1.66 $\pm$ 0.29 & 0.67 & \textbf{1.00} & 21.68 $\pm$ 4.05 \\
128 & 3 & silu & 1.66 $\pm$ 0.35 & 0.83 & \textbf{1.00} & 20.65 $\pm$ 4.18 \\
128 & 3 & softplus & 1.67 $\pm$ 0.17 & 1.35 & \textbf{1.00} & 20.77 $\pm$ 3.69 \\
128 & 3 & tanh & 2.45 $\pm$ 3.73 & 1.08 & 0.97 & 19.13 $\pm$ 3.78 \\
\hline
\end{tabular}
\caption{Architecture sweep for the K channel recovery task across 20 runs per configuration. Convergence is defined as RMSE $< 2$\,mV. A representative subset of configurations is shown; width $\geq 64$ reliably achieves full convergence regardless of activation function.}
\label{tab:arch_channel}
\end{table}

\paragraph{Runtime comparison.} Table~\ref{tab:runtime_channel} compares integration and forward-pass times for the hybrid model against a standard soma-only model across different network sizes. The overhead is moderate: for the default configuration (width 64, depth 2), integration time is comparable to the baseline, and the forward pass is approximately $6\times$ slower due to the neural network evaluation.

\begin{table}[h]
\centering
\begin{tabular}{cccccc}
\toprule
width & depth & $t_{\text{integrate}}$ & $t_{\text{forward}}$ & $\text{speedup}_{\text{integrate}}$ & $\text{speedup}_{\text{forward}}$ \\
\midrule
- & - & 0.081 $\pm$ 0.001 & \textbf{0.003 $\pm$ 0.001} & 1.000 & \textbf{1.000} \\
\hline
8 & 2 & 0.090 $\pm$ 0.002 & 0.017 $\pm$ 0.001 & 0.897 & 0.161 \\
8 & 3 & 0.093 $\pm$ 0.002 & 0.022 $\pm$ 0.000 & 0.865 & 0.122 \\
16 & 2 & \textbf{0.047 $\pm$ 0.005} & 0.017 $\pm$ 0.001 & \textbf{1.721} & 0.162 \\
16 & 3 & 0.054 $\pm$ 0.009 & 0.016 $\pm$ 0.001 & 1.505 & 0.165 \\
32 & 2 & 0.055 $\pm$ 0.008 & 0.017 $\pm$ 0.000 & 1.460 & 0.160 \\
32 & 3 & 0.065 $\pm$ 0.008 & 0.022 $\pm$ 0.001 & 1.238 & 0.121 \\
64 & 2 & 0.060 $\pm$ 0.009 & 0.017 $\pm$ 0.001 & 1.351 & 0.163 \\
64 & 3 & 0.072 $\pm$ 0.006 & 0.023 $\pm$ 0.000 & 1.126 & 0.118 \\
128 & 2 & 0.080 $\pm$ 0.007 & 0.017 $\pm$ 0.001 & 1.005 & 0.160 \\
128 & 3 & 0.115 $\pm$ 0.005 & 0.023 $\pm$ 0.000 & 0.705 & 0.119 \\
\bottomrule
\end{tabular}
\caption{Runtime comparison for the K channel hybrid model. Row 1 is the standard soma-only Hodgkin--Huxley model (baseline). Integration times for mid-sized networks are comparable to or faster than the baseline, likely due to the smoother dynamics learned by the neural ODE requiring fewer adaptive solver steps.}
\label{tab:runtime_channel}
\end{table}

\subsection{Learning Effective Axial Currents from Morphologically Detailed Neurons}
 
\begin{figure}[h]
    \centering
    \includegraphics[width=1.0\linewidth]{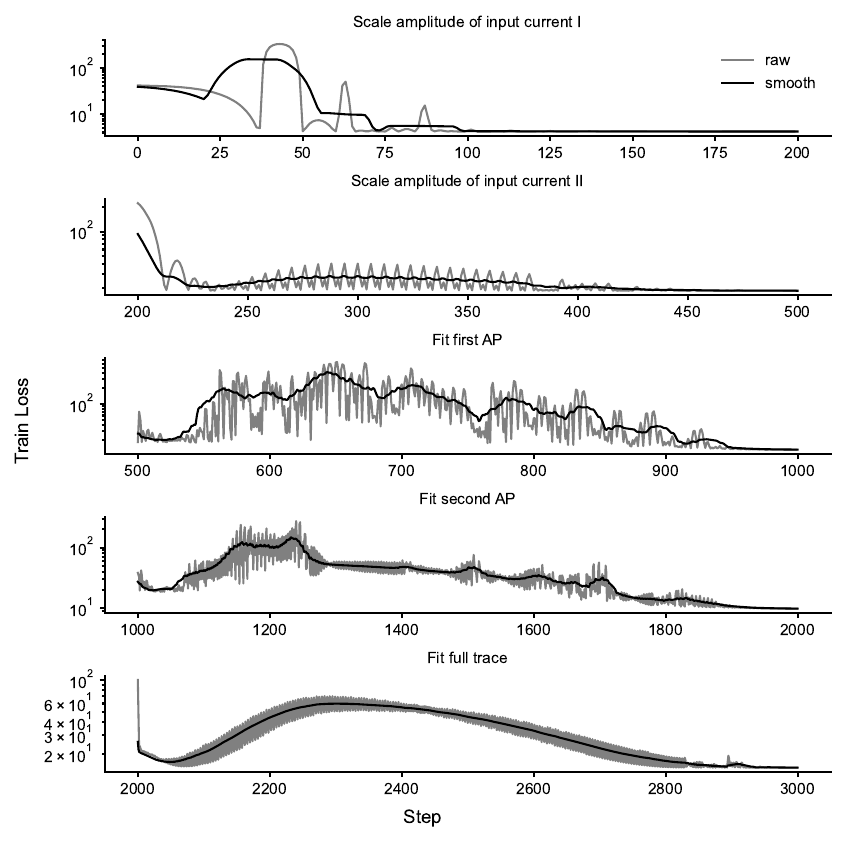}
    \caption{Training loss curves for the multicompartment reduction experiment (Sec.~\ref{sec:results_axial}). The five-stage curriculum proceeds as follows: (1--2)~scaling the input current amplitude, (3)~fitting the first action potential, (4)~fitting the second spike to reinforce the learned limit cycle, (5)~fitting the full trace. The oscillatory pattern within each stage reflects the cosine one-cycle learning rate schedule.}
    \label{fig:multicompartment_losses}
\end{figure}

\paragraph{Training dynamics.} The loss curves for the multicompartment fitting (Fig.~\ref{fig:multicompartment_losses}) show the five-stage curriculum. The model first learns to scale the amplitude of the input current to match the overall excitability, then fits the first action potential, followed by the second spike to reinforce the emerging limit cycle, before fitting the full trace. The oscillatory shape of the loss within each stage reflects the cosine one-cycle learning rate schedule.
 
\paragraph{Architecture sweep.} Table~\ref{tab:arch_multicomp} reports results across latent dimensions (1--8), network width (32--512), and depth (2--3). The formulation is robust across a wide range of configurations: even a single latent variable with a modest network (width 32, depth 2) achieves full convergence across 20 different runs. Performance degrades for overly large networks (width $\geq 256$ with depth 3, or width 512), likely due to optimization difficulties, or just not ever fully converging. Training times range from $\sim$5 minutes for the smallest configurations to over 20 hours for the largest.

\begin{table}[h]
\centering
\begin{tabular}{cccccccccc}
\toprule
$N_{z}$ & width & depth & RMSE & min. RMSE & frac. converged & runtime (hr) \\
\midrule
1 & 32 & 2 & 4.80 $\pm$ 0.11 & 4.59 & \textbf{1.00} & \textbf{0.19 $\pm$ 0.01} \\
1 & 32 & 3 & 5.08 $\pm$ 0.36 & 4.76 & 0.60 & 0.40 $\pm$ 0.42 \\
1 & 64 & 2 & 4.49 $\pm$ 0.11 & 4.27 & \textbf{1.00} & 0.23 $\pm$ 0.04 \\
1 & 64 & 3 & 4.57 $\pm$ 0.30 & 4.15 & 0.95 & 0.41 $\pm$ 0.21 \\
1 & 128 & 2 & 4.19 $\pm$ 0.15 & 3.85 & \textbf{1.00} & 0.31 $\pm$ 0.04 \\
1 & 128 & 3 & 4.07 $\pm$ 0.24 & 3.85 & \textbf{1.00} & 1.02 $\pm$ 0.69 \\
1 & 256 & 2 & 4.47 $\pm$ 0.90 & 3.70 & 0.80 & 0.93 $\pm$ 0.20 \\
1 & 256 & 3 & \textbf{4.00 $\pm$ 0.33} & 3.60 & \textbf{1.00} & 7.33 $\pm$ 5.63 \\
1 & 512 & 2 & 6.03 $\pm$ 5.57 & 3.38 & 0.60 & 7.56 $\pm$ 4.55 \\
1 & 512 & 3 & 7.43 $\pm$ 6.39 & 3.26 & 0.40 & 19.58 $\pm$ 9.38 \\
2 & 32 & 2 & 4.75 $\pm$ 0.09 & 4.59 & \textbf{1.00} & 0.19 $\pm$ 0.01 \\
2 & 32 & 3 & 4.86 $\pm$ 0.17 & 4.61 & 0.85 & 0.32 $\pm$ 0.06 \\
2 & 64 & 2 & 4.38 $\pm$ 0.12 & 4.13 & \textbf{1.00} & 0.24 $\pm$ 0.04 \\
2 & 64 & 3 & 4.65 $\pm$ 0.45 & 3.95 & 0.80 & 0.39 $\pm$ 0.12 \\
2 & 128 & 2 & 4.10 $\pm$ 0.14 & 3.82 & \textbf{1.00} & 0.36 $\pm$ 0.07 \\
2 & 128 & 3 & 4.08 $\pm$ 0.30 & 3.67 & \textbf{1.00} & 0.67 $\pm$ 0.40 \\
2 & 256 & 2 & 4.04 $\pm$ 0.29 & 3.72 & \textbf{1.00} & 1.10 $\pm$ 0.26 \\
2 & 256 & 3 & 4.95 $\pm$ 2.08 & 3.50 & 0.75 & 3.57 $\pm$ 2.52 \\
2 & 512 & 2 & 7.27 $\pm$ 6.84 & 3.36 & 0.45 & 6.89 $\pm$ 1.54 \\
2 & 512 & 3 & 6.16 $\pm$ 2.57 & 3.44 & 0.35 & 18.83 $\pm$ 8.32 \\
4 & 32 & 2 & 4.68 $\pm$ 0.15 & 4.24 & \textbf{1.00} & 0.21 $\pm$ 0.01 \\
4 & 32 & 3 & 4.68 $\pm$ 0.19 & 4.11 & \textbf{1.00} & 0.39 $\pm$ 0.14 \\
4 & 64 & 2 & 4.31 $\pm$ 0.25 & 3.64 & \textbf{1.00} & 0.27 $\pm$ 0.08 \\
4 & 64 & 3 & 4.44 $\pm$ 0.24 & 4.01 & 0.95 & 0.45 $\pm$ 0.20 \\
4 & 128 & 2 & 4.14 $\pm$ 0.50 & 3.40 & 0.95 & 0.33 $\pm$ 0.10 \\
4 & 128 & 3 & 4.21 $\pm$ 0.32 & 3.66 & 0.95 & 1.98 $\pm$ 6.17 \\
4 & 256 & 2 & 4.16 $\pm$ 0.57 & 3.56 & 0.95 & 1.30 $\pm$ 0.71 \\
4 & 256 & 3 & 4.07 $\pm$ 0.81 & \textbf{3.02} & 0.85 & 6.70 $\pm$ 6.92 \\
4 & 512 & 2 & 5.48 $\pm$ 1.14 & 4.02 & 0.45 & 5.93 $\pm$ 0.92 \\
4 & 512 & 3 & 7.66 $\pm$ 5.03 & 3.14 & 0.20 & 14.87 $\pm$ 5.05 \\
8 & 32 & 2 & 4.69 $\pm$ 0.13 & 4.43 & 0.95 & 0.25 $\pm$ 0.01 \\
8 & 32 & 3 & 4.70 $\pm$ 0.31 & 4.27 & 0.95 & 0.40 $\pm$ 0.06 \\
8 & 64 & 2 & 4.58 $\pm$ 0.29 & 4.10 & 0.95 & 0.29 $\pm$ 0.02 \\
8 & 64 & 3 & 4.67 $\pm$ 0.66 & 3.64 & 0.85 & 0.34 $\pm$ 0.05 \\
8 & 128 & 2 & 4.96 $\pm$ 0.85 & 4.09 & 0.70 & 0.33 $\pm$ 0.04 \\
8 & 128 & 3 & 4.75 $\pm$ 0.64 & 4.09 & 0.65 & 0.48 $\pm$ 0.09 \\
8 & 256 & 2 & 5.34 $\pm$ 0.81 & 3.88 & 0.40 & 1.29 $\pm$ 0.09 \\
8 & 256 & 3 & 5.26 $\pm$ 0.97 & 3.92 & 0.45 & 2.82 $\pm$ 0.41 \\
8 & 512 & 2 & 5.90 $\pm$ 0.76 & 3.75 & 0.10 & 4.62 $\pm$ 0.66 \\
8 & 512 & 3 & 8.28 $\pm$ 4.37 & 4.86 & 0.05 & 12.21 $\pm$ 4.57 \\
\bottomrule
\end{tabular}
\caption{Architecture sweep for the multicompartment reduction task across 20 runs per configuration. Convergence is defined as RMSE $< 5$\,mV. A representative subset is shown; the formulation is robust across a wide range of latent dimensions and network sizes, with degraded convergence only for overly large networks.}
\label{tab:arch_multicomp}
\end{table}

Note, that the RMSE for which solutions are considered converged is higher than for the potassium channel experiments, since we also use much longer simulation times in the axial current emulation experiments.

\paragraph{Runtime comparison.} Table~\ref{tab:runtime_multicomp} compares integration and forward-pass times for the hybrid model, the bare soma model, and the full 154-compartment simulation. The hybrid model achieves a substantial speedup over the full multicompartment model (up to $\sim$9.5$\times$ for the default hybrid configuration with width 32), while the overhead relative to the soma-only model remains moderate for small to mid-sized networks. For large networks (width $\geq 256$), the integration time approaches or exceeds that of the full multicompartment model, as the neural network evaluation dominates the computational cost.

\begin{table}[h]
\centering
\begin{tabular}{ccccccc}
\toprule
width & depth & $N_{comp}$ & $t_{\text{integrate}}$ & $t_{\text{forward}}$ & $\text{speedup}_{\text{integrate}}$ & $\text{speedup}_{\text{forward}}$ \\
\midrule
- & - & 1 & 0.047 $\pm$ 0.001 & 0.003 $\pm$ 0.001 & 23.782 & 9.422 \\
- & - & 154 & 1.110 $\pm$ 0.044 & 0.032 $\pm$ 0.000 & 1.000 & 1.000 \\
\midrule
32 & 2 & 1 & \textbf{0.117 $\pm$ 0.010} & \textbf{0.008 $\pm$ 0.001} & \textbf{9.481} & 3.990 \\
32 & 3 & 1 & 0.130 $\pm$ 0.010 & 0.009 $\pm$ 0.001 & 8.541 & 3.460 \\
64 & 2 & 1 & 0.132 $\pm$ 0.008 & \textbf{0.008 $\pm$ 0.001} & 8.425 & \textbf{4.029} \\
64 & 3 & 1 & 0.167 $\pm$ 0.008 & 0.009 $\pm$ 0.001 & 6.653 & 3.478 \\
128 & 2 & 1 & 0.184 $\pm$ 0.009 & \textbf{0.008 $\pm$ 0.001} & 6.020 & 3.992 \\
128 & 3 & 1 & 0.267 $\pm$ 0.011 & 0.009 $\pm$ 0.000 & 4.163 & 3.449 \\
256 & 2 & 1 & 0.673 $\pm$ 0.019 & \textbf{0.008 $\pm$ 0.001} & 1.649 & 3.956 \\
256 & 3 & 1 & 1.556 $\pm$ 0.058 & 0.009 $\pm$ 0.000 & 0.713 & 3.507 \\
512 & 2 & 1 & 2.629 $\pm$ 0.066 & \textbf{0.008 $\pm$ 0.001} & 0.422 & 3.903 \\
512 & 3 & 1 & 4.913 $\pm$ 0.188 & 0.010 $\pm$ 0.000 & 0.226 & 3.339 \\
\bottomrule
\end{tabular}
\caption{Runtime comparison for the multicompartment reduction task. Speedups are relative to the full 154-compartment simulation. The runtimes are averaged across models with 1, 2, 4 and 8 latent states, as comparably the number of latent dimensions did not have a big impact on the runtimes as opposed to other factors like network width. The hybrid model with moderate network sizes (width 32--128) achieves 6--10$\times$ speedup over the full model. For width $\geq 256$, the neural network evaluation dominates and the speedup diminishes.}
\label{tab:runtime_multicomp}
\end{table}

\newpage
\clearpage

\renewcommand{\thesection}{\arabic{section}}
\renewcommand{\thefigure}{\arabic{figure}}
\renewcommand{\thetable}{\arabic{table}}
\renewcommand{\theequation}{\arabic{equation}}



\end{document}